# Lung Nodule Classification Using Biomarkers, Volumetric Radiomics and 3D CNNs

*Kushal Mehta; Arshita Jain; Jayalakshmi Mangalagiri; Sumeet Menon; Phuong Nguyen, PhD; David R. Chapman, PhD;*
*University of Maryland Baltimore County*

*Abstract*— **We present a hybrid algorithm to estimate lung nodule malignancy that combines imaging biomarkers from Radiologist's annotation with image classification of CT scans. Our algorithm employs a 3D Convolutional Neural Network (CNN) as well as a Random Forest in order to combine CT imagery with biomarker annotation and volumetric radiomic features. We analyze and compare the performance of the algorithm using only imagery, only biomarkers, combined imagery + biomarkers, combined imagery + volumetric radiomic features and finally the combination of imagery + biomarkers + volumetric features in order to classify the suspicion level of nodule malignancy. The National Cancer Institute (NCI) Lung Image Database Consortium (LIDC) IDRI dataset is used to train and evaluate the classification task. We show that the incorporation of semi-supervised learning by means of K-Nearest-Neighbors (KNN) can increase the available training sample size of the LIDC-IDRI thereby further improving the accuracy of malignancy estimation of most of the models tested although there is no significant improvement with the use of KNN semi-supervised learning if image classification with CNNs and volumetric features are combined with descriptive biomarkers. Unexpectedly, we also show that a model using image biomarkers alone is more accurate than one that combines biomarkers with volumetric radiomics, 3D CNNs, and semi-supervised learning. We discuss the possibility that this result may be influenced by cognitive bias in LIDC-IDRI because malignancy estimates were recorded by the same radiologist panel as biomarkers, as well as future work to incorporate pathology information over a subset of study participants.**

*Keywords—Lung cancer, LIDC, Malignancy, Benign, Radiomics, Biomarkers, CNNs.*

I. INTRODUCTION

Lung cancer accounts for the highest number of cancer related deaths globally, but early detection can improve prognosis. Lung cancer screening using Low Dose Computed Tomography (LDCT) has become standard practice as a way of determining which pulmonary nodules are likely benign and which nodules require biopsy to determine malignancy. However, lung cancer screening has a high false positive rate clinically due to the need to identify a large percentage of malignant nodules for biopsy. Thus, many biopsies are performed on patients that ultimately do not have cancer. Our goal is to develop a hybrid Computer Aided Diagnosis (CAD) algorithm that combines CNN based image classification, with volumetric Radiomics, as well as descriptive biomarkers from Radiologists annotation. We also evaluate the extent to which descriptive biomarkers can be used for the purposes of semi-supervised learning in order to help to reduce this false positive rate.

The main objective of this paper is to determine to what extent that imaging biomarkers (from radiologist annotation) can be combined with automated 3D image classification of CT scans in order to create a hybrid human knowledge + machine learning algorithm for the assessment of lung nodule malignancy. In this paper, we present an algorithm for classifying the lung nodule malignancy suspicion level as either being malignant or benign where malignant means that the nodule is highly suspicious and benign being that the nodule is highly not suspicious. Our approach is to combine these data sources using a CNN as well as a random forest classification algorithm.

The novelty of our approach lies in two factors that have not been previously explored utilizing the LIDC-IDRI dataset for lung nodule malignancy estimation. #1 The integration and impact analysis of the combination of image classification (CNN+volumetric radiomics) in tandem with descriptive biomarkers from radiologists annotation. #2 The use of descriptive biomarkers as a foundation for semi-supervised learning in order to make use of nodules with "intermediate" malignancy as part of the training data. We describe as part of our literature review, that these two techniques have not yet been analyzed to determine their relative impacts on lung nodule malignancy estimation. As such, our approach implements both of these methodologies and we make use of hypothesis testing in order to determine the extent to which combination either or both of these techniques improves the sensitivity and specificity of malignancy estimation as measured by AUC. The goal is to show that both of these techniques have the ability to increase the AUC of lung nodule malignancy estimation as compared to baseline algorithms that do not take these factors into account.

For the classification of lung nodule malignancy, we use the scores of levels 1 to 5 which are labeled by the four board-certified radiologists for the creation of the LIDC-IDRI dataset for lung nodules >=3mm. These scores range from 1 to 5 with 1 meaning highly unlikely to be malignant, 2 as moderately unlikely, 3 as intermediate, 4 as moderately suspicious to be malignant and 5 as highly likely to be malignant. The total number of lung nodules after removing inconsistent data is 4505.

We assess and compare the classification accuracy of malignancy suspicion by using general categories of computational strategies: radiologist identified image biomarkers using Random Forest, image classification using 3D CNNs, and hybrid algorithms that combine image features, biomarkers and volumetric radiomic features. In our assessment we will determine the discriminating power of each of these techniques individually as well as combined and also measure the impact of semi-supervised learning in improving the classification accuracy of the random forest algorithm.

An unexpected result of our analysis is that the biomarker only predictions perform better and achieve higher AUC than the hybrid model that combines biomarkers, volumetric radiomics and 3D CNNs. This would suggest that descriptive biomarkers are superior to volumetric radiomics as well as 3D CNNs. However, it is impossible to eliminate the possibility of cognitive bias in the LIDC-IDRI, because the biomarkers are recorded by the same radiologist panel that estimated malignancy [14].

A. *What are Biomarkers*

The National Institutes of Health biomarkers definitions working group defined a biomarker as "a characteristic that is objectively measured and evaluated as an indicator of normal biological processes, pathogenic processes, or pharmacologic response to a therapeutic intervention."

In the area of medical imaging, an imaging biomarker is a characteristic feature of an image which is applicable to a patient's diagnosis. Imaging biomarkers play a vital role in major medical fields such as oncology as they are widely useful in predicting the lung nodule malignancy suspicion. There are several applications of biomarkers which includes prediction, detection, staging, grading and evaluation of responsiveness to the treatment. Imaging biomarkers are utilized for all of these applications and they have the advantage of being non-invasive and being spatially and temporally resolved. The biomarkers of a nodule independent on each of the CT scans include subtlety, internal structure, calcification, sphericity, margin, lobulation, spiculation, texture and malignancy. Each of these characteristic features are explained briefly in the following sections

B. *What are volumetric radiomic features?*

Lung CT scans contain features that are either qualitative or quantitative. These features reflect the pathophysiology of the nodule. These quantitative features are extracted from the image with the help of mathematical and data characterization algorithms. This process is known as radiomics and the extracted quantitative features are known as radiomic features. V. Parekh et. al [9] define it as, 'Radiomics is the high throughput extraction of quantitative features from radiological images creating a high dimensional data set followed by data mining for potentially improved decision support'. The radiomic features mainly comprise of texture, shape and gray level statistics of the nodule. We focus on the shape and volumetric features in this study. Specifically, the maximum diameter of the nodule, the surface area and the volume of the nodule. The units for each of these features are in mm, $mm^2$ and $mm^3$.

II. LITERATURE REVIEW

Algorithms and statistical methodologies to predict the probability of lung nodule malignancy are an area of active research. Many related papers have developed methodologies to estimate the probability of malignancy using either human annotated imaging biomarkers, or automated image classification algorithms. However, few papers have attempted to combine this human annotation with automated image classification into a single technique.

The analysis of imaging biomarkers and their discriminating power for nodule malignancy is an important area of research. Liu et al. (2017) [1] present a cross-validated analysis where they identify the radiological image characteristics most helpful in predicting the risk of malignancy in lung nodules. These semantic qualities are further integrated with size-based measures which leads to enhanced prediction of accuracy. Their methodology measures the quantity of incidentally recognized pulmonary nodules depending on some observed radiological features measured on a point scale and followed by a machine-learning method which utilizes this information in predicting the status of cancer. Hancock & Magnan (2016) [2], explored the predictive ability of statistical learning methods for classifying the malignancy of lung nodules using LIDC dataset and the radiologist annotations for the lung nodules from where they derived estimates of the diameter and volume of the lung nodules. Also, the paper strongly states that the lung nodules can be classified as malignant or benign by just using quantified, diagnostic image features. In their classification of malignancy, they have analyzed how accurately the malignancy could be classified depending upon the particular features and feature subsets and have ranked spiculation, lobulation, subtlety and calcification to be the top four features that is having higher predictive power than others. They have calculated theoretical upper bounds on the accuracy of classification which can be achieved by an ideal classifier by just using the annotated feature values assigned by the radiologist which can get can accuracy of 85.74% which is 4.43% below the theoretical maximum of 90.17% with AUC score of 0.932. But when they have also considered diameter and volume features then this AUC score has enhanced to 0.949 with an accuracy of 88.08%. As such it is clear that machine learning using radiologist annotated imaging biomarkers is a viable approach.

A separate branch of related research studies has investigated automated image classification techniques. In recent years two main branches of image classification methods have been investigated (a) automated feature extraction using deep neural networks (b) hand-crafted feature extraction using radiomics. Deep neural networks attempt to learn a representation and have outperformed the use of hand-crafted features on general image classification tasks [3] [4] [5]. Radiomic however are quantifiable radiological features that can be algorithmically extracted through hand-crafted computer vision techniques.

The use of deep neural networks was first explored by Kumar, D., Wong, A., & Clausi, D. A. (2015, June) [6] the authors applied deep autoencoders with a binary decision tree classifier to estimate malignancy of 4323 nodules of the LIDC-IDRI dataset, achieving an overall accuracy of 75.01% with 83.35% sensitivity and a false positive of 0.39 per patient over a 10-fold cross validation.

3D CNN models for nodule malignancy were first explored by Li, W., Cao, P., Zhao, D., & Wang, J. (2016) [7]. The authors designed specific network architectures for three types of nodules namely solid, semisolid, and ground glass opacity (GGO). These architectures were trained using 62,492 regions-of-interest (ROIs)samples which includes 40,772 nodules and 21,720 non-nodules from LIDC-IDRI database. Furthermore, the paper has a deep CNN presented which is built on $32 * 32$ image ROI data. The need for separate architectures was resolved by Shen et. al (2017) [KM1] which presents a novel Multi-crop Convolutional Neural Network (MC-CNN). The MC-CNN model follows a novel multi-crop pooling strategy and has the added advantage of deriving nodule semantic attributes and diameter in addition to estimating the lung nodule malignancy. Apart from classifying the nodule malignancy suspicion, the authors in this paper have extended their proposed approach furthermore to evaluate uncertainty of malignancy by quantifying nodule semantic labels prediction which includes the characteristic features like subtlety and margin along with estimating the diameter of the nodule using their proposed multi-crop convolutional neural networks. This approach helps the researchers in the assessment of uncertainty of malignancy in lung nodules and their results in this paper seems to be motivating.

Zhao et. Al [5] employs a CNN called Agile CNN that uses a unique combination of LeNet [8] and Alexnet [9] for classification of the lung nodules in the LIDC dataset into malignant and benign class. The LeNet architecture is used with the parameter settings of Alexnet [9]. Inputs to this model are 2D images of shape 53 by 53 that contain non-centralized nodules. They achieve an accuracy of 82.2% and an AUC of 0.877 for a dataset that contains 243 CT image nodule samples. While traditional 2D CNN's are efficient in image classification problems, a CNN that takes 3D images as input is better suited to solve medical image classification problems. A 3D region of interest has the potential to encapsulate the entire nodule, and the CNN can extract abstract features in 3 dimensions.

Ensemble CNN estimation was investigated by Liao et al. (2019) [10] Their model has two modules where a 3-d region proposal network (RPN) detect suspicious nodules and the second module picks up the top five nodules which seems to be having more probability based on its evaluation and integrates them with a leaky noisy-OR gate for acquiring the probability of lung cancer in a patient. They used a modified U-net as their backbone network for both of its modules. All of these algorithms that employ deep neural networks have attempted to perform automated feature extraction, although recent methods have required architectural enhancements in order to extract features at a variety of scales. Our research differs from these approaches because we attempt to incorporate imaging biomarkers from human annotation in addition to automated feature extraction using 3D CNNs.

Radiomics based approaches have the advantage that the extracted features are quantifiable and have an intuitive radiological definition [15]. Rodrigues et. al (2018) [11] the authors aimed at classifying the lung nodules of being malignant or benign along with classifying the levels of malignancy (1-5). For such a classification, a novel structural co-occurrence matrix (SCM) dependent approach is used for the extraction of features though the images of nodules before classifying them as being malignant or benign along with the levels of malignancy. For this classification, the authors in this research have made use of three classifiers namely multilayer perceptron, support vector machine and k-nearest neighbors' algorithm.

The most similar approaches to our research are algorithms that combine multiple techniques. Causey et al. (2018) [12] have developed a highly accurate image classification algorithm named NoduleX which uses a combination of CNN as well as radiomics for prediction on lung nodule malignancy using CT images of LIDC data set. They also presented several variations of the radiomics as well as deep learning CNN hyperparameters. This paper detects high accuracy in classifying lung nodule malignancy with an AUC of 0.99. Our research differs from Causey et. al (2018) [12] in that we investigate the classification accuracy of 3D CNNs + imaging biomarkers (from radiologist annotation), as opposed to 3D CNNs + Radiomics.

In Li et al. (2019) [13] a fusion algorithm is proposed where they have integrated the highest level CNN representation learned at the output layer of a 3D CNN with the handcrafted features by utilizing the support vector machine that couples with the sequential forward features selection approach for selecting the optimal feature subset thereby building the final classifier. The authors of this paper claim that their fusion algorithm could pave a path to enhanced performance in recognizing the malignant and benign lung nodules by using the LDCT lung cancer screening dataset with a high sensitivity and specificity.

### III. MATERIALS AND DATASETS

#### A. LIDC-IDRI Dataset Overview

The LIDC-IDRI Dataset [8] is a publicly available dataset that consists of diagnostic and lung cancer screening thoracic computed tomography (CT) scans with marked-up annotated lesions. This dataset is a web-accessible international resource initiated by the National Cancer Institute (NCI), and then further developed by the Foundation for the National Institutes of Health (FNIH) and going along with the Food and Drug Administration (FDA). LIDC is used for the purpose of research towards development, training and assessing of computer-aided diagnostic (CAD) methods for detecting and diagnosing lung cancer in its early stages. This dataset is created in collaboration with seven academic centers and eight medical imaging companies that have 1018 CT cases where it has thoracic CT scan images as DICOM files, and an additional XML file, for each patient. The LIDC study has annotations that are provided by four experienced thoracic radiologists who reviewed each of all the 1018 CT cases in the LIDC/IDRI cohort and marked lesions into 3 categories based on the nodule size. The nodule > or= 3mm means that they have a greater probability of being malignant than lesions that have a nodule<3mm and non-nodule> or= 3mm. The malignancy rating is given from 1-5 depending on the size of the nodule.

#### B. Image Biomarkers

Each nodule >= 3mm in the LIDC-IDRI dataset contains the following image characteristics which were annotated by four independent board-certified radiologists,
- Subtlety: Difference between the lung and its surroundings.
- Internal Structure: Composition present internally in the nodule.
- Calcification: Appearance of calcium in the nodule. Smaller the nodule, the more likely it comprises calcium for visualization. A nodule being benign is highly related with a calcification rating that is central, non-central, laminated, and popcorn.
- Sphericity: The measurement of the shape of a nodule in terms of roundness.
- Margin: Looking for how well the margins of the nodule are established.
- Lobulation: If the lobular shape is clearly visible from the margin or not. If it is apparent, then it is a sign of a nodule being benign.
- Spiculation: Level to which a nodule is exhibiting spicules or spike-like formations along the border of lung nodule where spiculated margin indicates that the nodule is malignant.
- Texture: The density of the nodule internally which serves as an important characteristic when segmenting a nodule as partly solid and nonsolid texture can complicate when establishing the nodule boundary.
- Malignancy: Likelihood of growing cancer where malignancy is highly likely to be related with large nodule size and smaller nodules being more likely to be benign. Also, the nodules associated in being benign are noncalcified and include spiculated margins.

According to the LIDC/IDRI dataset, the classification of these biomarkers belongs to only the nodules whose size is >= 3mm and our model has trained accordingly for predicting the malignancy levels. Since our dataset has 1010 patients and nodules greater than the number of patients, we have found that 6859 nodules are >= 3mm according to annotations that are provided by the four experienced Radiologists. However, for a subset of 100 cases among the initial 399 cases released, there is some rate of inconsistency with respect to the spiculation and lobulation characteristics of lesions that are identified to be nodules> 3mm. Due to this inconsistency in the data, XML nodule characteristics data will have an impact and thus, we have excluded this inconsistent data by considering the CT scans that were present in the initial release. Therefore, we ended up having 4505 nodules which are >= 3mm.Each of these nodules now have 8 characteristics in total along with 1 characteristic called malignancy that is treated as a label for that particular nodule.

#### C. Nodule visualization

LIDC-IDRI consists of lung cancer screening thoracic computed tomography (CT) scans with marked-up annotated lesions. Each subject includes CT scan images and an associated XML file that records the results of image annotations by experienced thoracic radiologists. Each XML file consists of nodules divided in three broad categories: Nodules ≥ 3mm diameter, Nodules < 3mm diameter and Non-nodules ≥ 3mm diameter. For each nodule ≥ 3mm, each radiologist drew a complete outline around the nodules in all sections that it appeared, with the pixels that comprise the outline at the first pixel outside the nodule. These annotations are given in the form of nodule regions of interest and their z-positions. We used these spatial coordinates to construct a 3-D box and 3-D mask, centered at annotated locations of lung nodules, of a fixed size. We use box size of
32 pixel x 32 pixel x 16 slices for our experiments.

We combine the biomarkers, 3D image volume and the 3D image mask into a unified dataset that we use for all our experiments in this paper. The dataset is saved as a numpy array and contains the following information for every annotation in the LIDC Dataset.
- Subtlety
- Internal Structure
- Calcification
- Sphericity
- Margin
- Lobulation
- Speculation
- Texture
- Malignancy
- Patient ID
- 3D Volume
- 3D Mask

IV. APPROACH

We treat the Malignancy category approximation as a binary classification problem for 'Malignant' versus 'Benign' nodules by splitting the annotated nodules on the radiologist assigned malignancy scores. Note that the terms 'Malignant; and 'Benign' refer to the suspicion levels of the nodules as annotated by the radiologists in the LIDC dataset and not the actual malignancy level of the nodule. The class 'Benign' means that the nodule is highly unlikely to be malignant and the class 'Malignant' means the nodule is highly suspicious to be cancerous. These scores range from 1-5 (1 meaning highly unlikely to be malignant, 2 as moderately unlikely, 3 as intermediate, 4 as moderately suspicious to be malignant and 5 as highly suspicious to be malignant). We identify each of these sets as R1, R2, R3, R4 and R5. We show the number of nodules belonging to each of these classes in **Figure 1**.

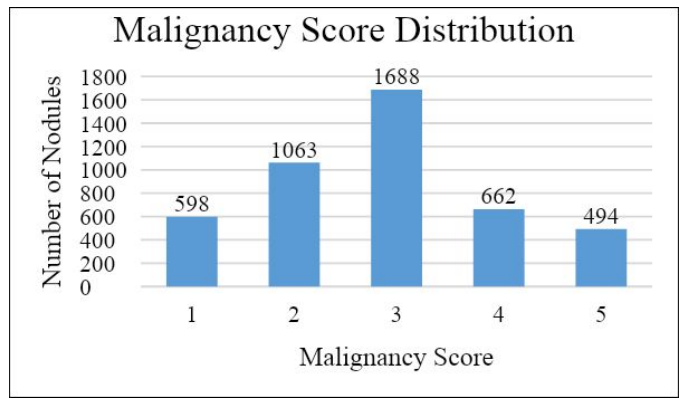

*Figure 1. – Number of nodules per malignancy level of the 4505 nodules in the dataset.*

As we perform binary classification, we must group the nodules based on their malignancy scores. The criteria to group the nodules into these classes differ for each distribution in the dataset that are described as follows:

- Distribution A (Fully Supervised Method) - R12 vs R45 – Nodules belonging to set R1 and R2 are grouped in a set we call R12 and are all labeled as class 0 ('Benign'). Similarly, set R45 consists of nodules that belong in the sets R4 and R5 and are given the label 1 ('Malignant'). This follows the approach of Hancock et. al (2016). Nodules labeled by a radiologist as having an intermediate malignancy (R3) are not considered for classification in this distribution. We have 2817 nodules after grouping the nodules and removing R3. Models that make use of this distribution for training and testing follow a fully supervised method. We shall refer to those models as fully supervised models throughout the paper.

- Distribution B (Semi-Supervised Method) - R123 vs R345 – The nodules belonging in the R3 set make up a 37.4% of the dataset that cannot be ignored as they lie close to the decision boundary. These nodules show biomarker similarities to nodules belonging in R12 and R45. We use a K-Nearest Neighbors algorithm to identify nodules that have closest biomarker features to those nodules that belong in the 'Malignant' and 'Benign' class. These nodules are then identified according to their closeness. The value of K is empirically determined to be 21. The distance metric is the Euclidean distance. Nodules belonging to set R1, R2 and a subset of nodules in set R3 are grouped in a set R123 and are labeled as 0 ('Benign'). R345 consists of the remaining nodules in the set R3, and all nodules in R4 and R5. This superset is given the label of 1 ('Malignant').

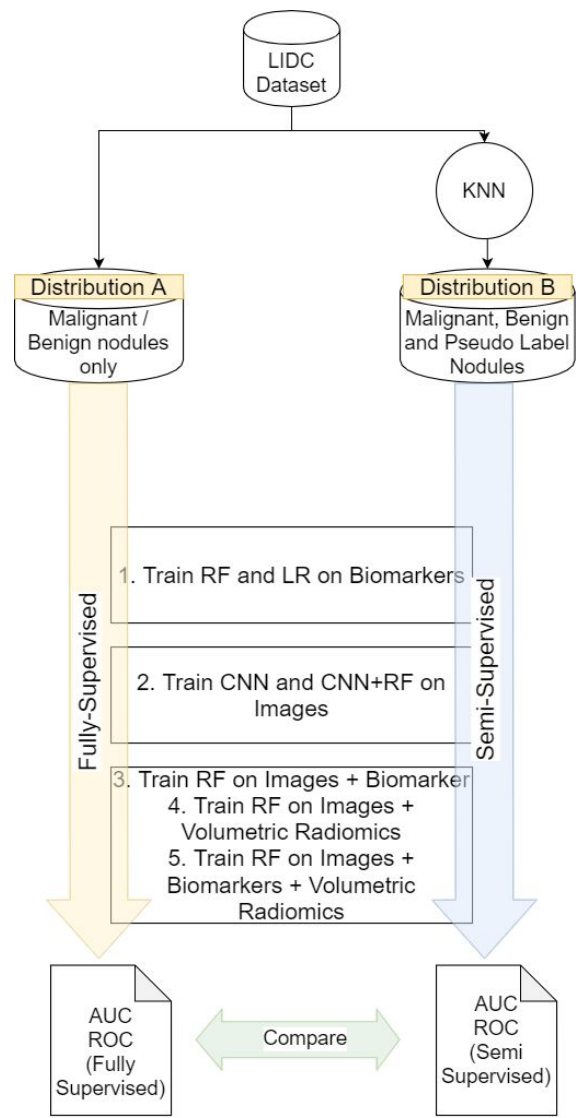

*Figure 2. Overview of all the approaches taken in all experiments.*

As illustrated in **Figure 2,** features representing the nodules are used to train various models in two branches. Both branches consist of the experiments to train the following -

1. A random forest and logistic regression model on biomarkers only
2. A 3D CNN on nodule images and a Random Forest on the deep image features extracted from the 3D CNN.
3. A Random Forest on combination of deep image features and biomarkers.
4. A Random Forest on combination of deep image features and volumetric radiomics.
5. A Random Forest on combination of deep image features, biomarkers and volumetric radiomics.

The first branch on the left performs these experiments using the fully-supervised method. The second branch performs these experiments using the semi-supervised method. These two branches are independent and each experiment results in a ROC AUC value that is used for comparison between the supervised and semi-supervised methods.

### Steps to find best K - Illustrated in Figure 3
1. Create data Distribution A by grouping the R1, R2 nodules and R4, R5 nodules. Hold out the nodules that belong to R3.
2. Split Distribution A into a train and validation set with a random 80:20 split.
3. Train and validate KNN with values of K ranging from 1 to 51 in steps of 2. In simple terms, we select a value of K that is an odd number in the 1-51 range and train the KNN on the training set with that K value. We evaluate the trained KNN model on the validation dataset and record it. Totally we have 26 values of K for which the KNN is trained and validated. We select the value of K that gives the best validation accuracy.
4. Steps 2 and 3 are repeated 1000 times, and the train and validation accuracy are recorded for each run. We end up with 1000 best validation accuracies and the corresponding K values that led to that accuracy.
5. The value of K that has the highest accuracy and also the highest frequency in the 1000 runs is selected as the optimal K value for the KNN.

All the experiments follow two classification techniques –

1. **Fully Supervised** – When a model employs this technique, the train set, and test set are derived from data distribution A with a random seed to perform the split. Both the train and test sets include nodules that belong to either R12 or R45. Nodules that have a malignancy of 3 are not included in this dataset.
2. **Semi Supervised** – When a model employs a semi-supervised strategy, a train and test set is derived from distribution B with a random seed for splitting. The R3 nodules in the train set undergo a KNN pseudo-labeling step to classify them into the Benign (R12) or Malignant (R45) class based on the euclidean distance of descriptive biomarkers. Since a portion of the R3 nodules in the train set get classified into the benign class and the remaining get classified into the malignant class, we refer to that split as R123 vs R345. The R3 nodules are removed from the test set in order to maintain consistency and fairness while comparing the model's performance with the fully supervised technique.

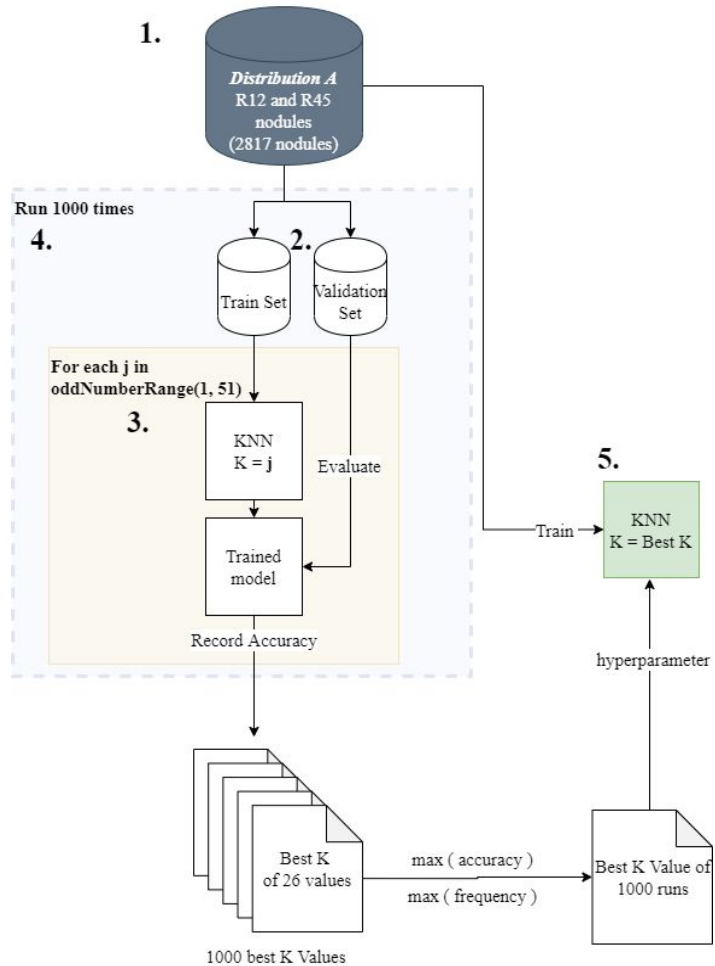

*Figure 3. KNN - Best K Selection Process*

*Convolutional Neural Network Architecture*

For experiments involving CNN image classification, we make use of the following CNN model as shown in **Figure 4**, the 3-D Lung volume is passed through the first layer of CNN with the filter of size (3*3*3) and the output of the first layer is of the shape (32,32,16,32). It is then passed on to the next layer of convolution which gives us an output of the shape (32,32,16,32). The kernel size for all the convolutional layers are (3,3,3). The regularizers are used with a regularization factor of 0.01. Also, activation function after each conv layer is relu and the maxpool layer has a pool size of (2,2,2). After the 2nd layer of convolution, it is then passed on to the 3rd layer for max-pooling which gives us an output of the shape (16,16,8,32). After the image is down-sampled in the max-pooling layer it is then passed on to 2 more convolution layers which make up the 4th and 5th layer of the model. This 4th layer gives us output of shape (16,16,8,64) and the 5th layer gives us an output of (16,16,8,64) which is then passed on to the next layer for max-pooling. The next max-pooling layer returns a down-sampled output of (8,8,4,64). The next convolution layer takes the down-sampled image as input and returns the output image of shape (8,8,4,128) and is then passed on to the next convolution layer which returns an output of shape (4,4,2,128) and is passed on to the next max-pooling layer. This max-pooling layer (Layer 9) returns an output of shape (8,8,4,128) and is then passed on to another set of two convolution layers. The 10th convolution layer further performs convolution over this image and returns an output of shape (4,4,2,256) which is passed on to the 11th convolution layer as input and further returns an output image of shape (4,4,2,256).

This same action is later performed on another set of 1 max-pooling layer and 2 consecutive convolution layers. The 12th layer which is the max-pooling layer performs down-sampling on the image which return an output of shape (2,2,1,256) and is fed as input to the 13th layer which performs convolution, passes it to the 14th layer for convolution and finally returns an output of shape (2,2,1,512). The final 3 layers are fully-connected dense layers with 1024 neurons, 64 neurons and 1 neuron respectively.

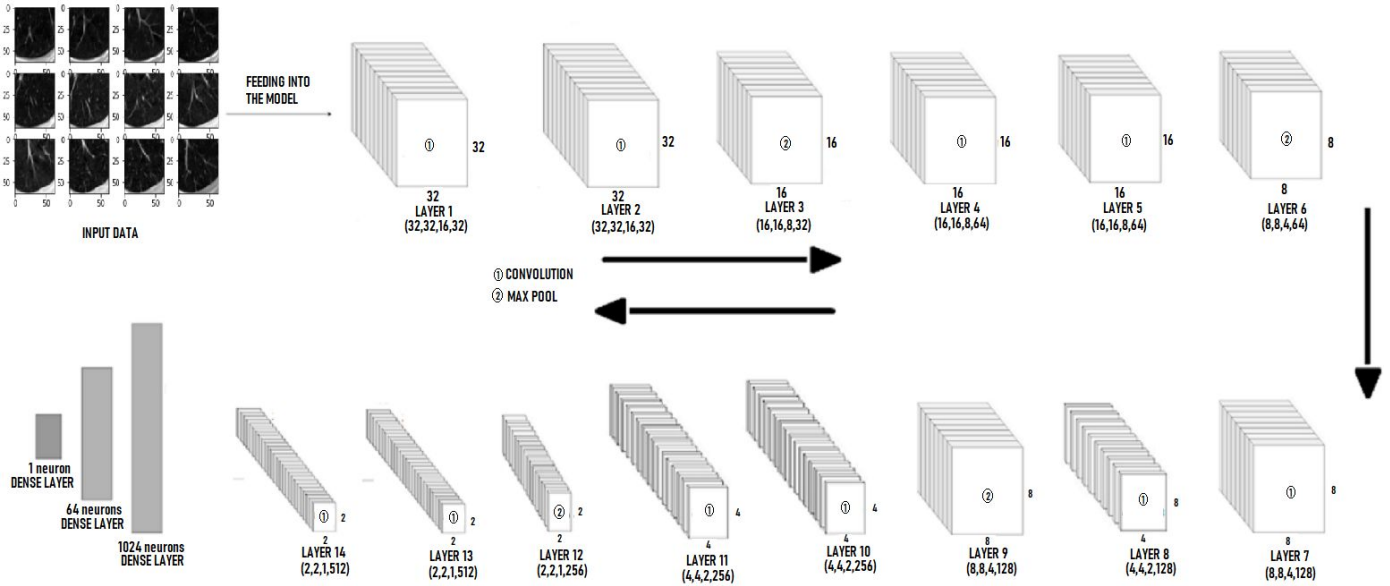

Figure 4. Malignet

## V. EXPERIMENTAL DESIGN

We perform 5 experiments in order to evaluate the performance of the combination of nodule malignancy classification using CNNs, biomarkers, and volumetric radiomics as follows,
1. Experiment 1 - Classification using Biomarkers only
2. Experiment 2 - Classification using Images only
3. Experiment 3 - Classification using Images + Biomarkers
4. Experiment 4 - Classification using Images + Volumetric Radiomics
5. Experiment 5 - Classification using Images + Biomarkers + Volumetric Radiomics

Our evaluation criteria is the Area Under Curve (AUC) of the Receiver Operating Curve (ROC). For each model ROC curves are presented by varying the prediction threshold between 0 to 1 and plotting sensitivity against 1 - specificity. AUC varies from 0 to 1 (higher is better), and is defined as the integral of sensitivity with respect to 1-specificity over the domain of the ROC curve. For each comparison of models we calculate p-value of the difference in AUC using a two-tailed Student's t-test in order to determine statistical significance of one classification model versus another. We consider this difference to be significant if the p-value is smaller than our α-value of 0.05.

### A. *Experiment 1 – Classification using Biomarkers only*

A logistic regression and a random forest model are trained using the fully supervised method. The train and test set are derived from distribution A. The split ratio is 80:20. The grouped malignancy suspicion score (R12 'Benign' & R45 'Malignant') are taken as labels.

An additional set of logistic regression and random forest models are trained using the semi-supervised method. The train and test sets are derived from distribution B. The split ratio here is 80:20 as well. The R3 nodules in the train set are given pseudo labels as 'malignant' or 'benign' using KNN. The train set contains nodules from R12 and pseudo labels from R3, which forms the R123 group with the label 'Benign' and nodules from R45 and the remainder of the pseudo-labeled R3 group that forms R345 with the label 'Malignant' as described above. The R3 nodules are removed from the test set explicitly.

These models are trained on the biomarker features only. These are subtlety, internal structure, calcification, sphericity, margin, lobulation, spiculation and texture. We measure the Receiver operating characteristic area under the curve (ROC AUC) on the test set for the fully supervised models and the semi supervised models independently. The process of creating train and test sets and training all the models using both type of supervision methods is repeated 1000 times for robust results. The AUC values are recorded for every iteration on both models and the average AUC is calculated and the average ROC is plotted. **Figure 5** illustrates the entire process for both methods completely.

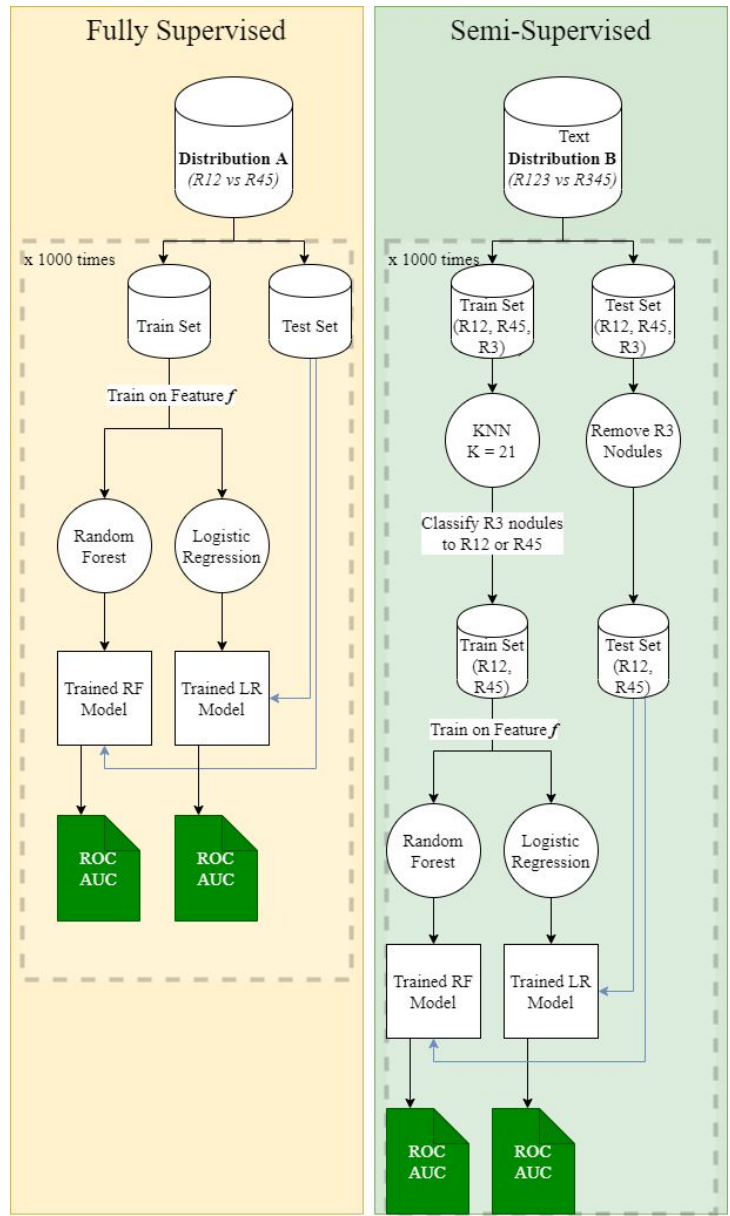

*Figure 5. Training and testing process for the fully supervised and semi-supervised method for experiment 1. Feature f here refers to biomarkers.*

B. *Experiment 2 – Classification using Images only*

We employ a ten layer 3D CNN model and a 3D CNN + Random Forest fusion model using the fully supervised method and the semi-supervised method independently. For the fully supervised method, the train and test sets are created from distribution A with an 80:20 split as done in the previous experiments. 3D Bounding Boxes that encapsulate the voxels representing the nodule in the center of the box are used to train the 10 layer 3D CNN. The bounding boxes from the test set are then used to obtain the predictions from the trained model. Using these predictions, the ROC AUC value is calculated. We use this trained model for training the 3D CNN + RF fusion model.

We extract the feature vectors of the bounding boxes from the second to last dense layer for each row in the train and test set. These are single dimension feature vectors with a length of 64. Using the reference of the train set and test set, we assign the original labels from the dataset to the feature vectors and obtain a train and test set for the random forest. After training the random forest on these feature vectors, we calculate the ROC AUC values and plot them.

The same approach is applied when the semi-supervised method is used. Train and test sets are derived from distribution B with a random split of 80:20. The only difference is that KNN with K=21 is used to classify the R3 nodules from the train set into benign and malignant classes. R3 nodules are dropped from the test set.

The fully supervised and semi-supervised implementations of the 3D CNN and 3D CNN + Random forest models are trained and tested 30 times to obtain robust results. The number 30 is decided as a sufficient number of iterations to calculate statistical significance between the average AUCs between models. This decision is also attributed to the limitations in computing power and the amount of time one CNN model takes to train. The trained CNN model, train and test sets for both the supervised and semi-supervised method are saved in each iteration. This experiment design is illustrated in **Figure 6**.

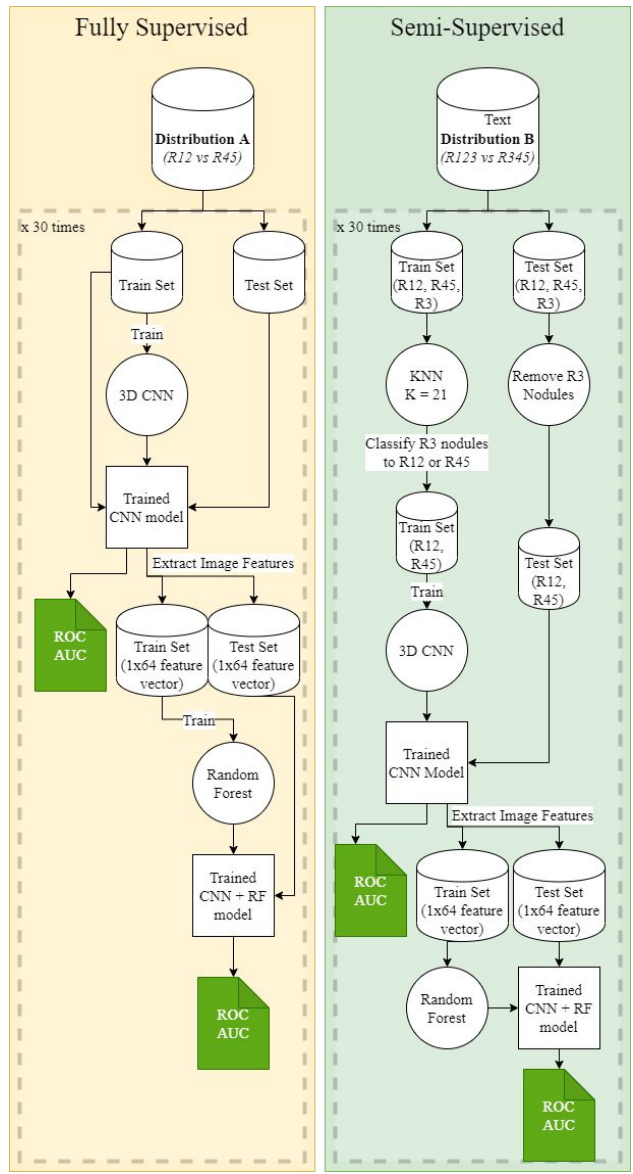

*Figure 6. Training and testing strategy for the fully supervised and semi-supervised methods for experiment 2.*

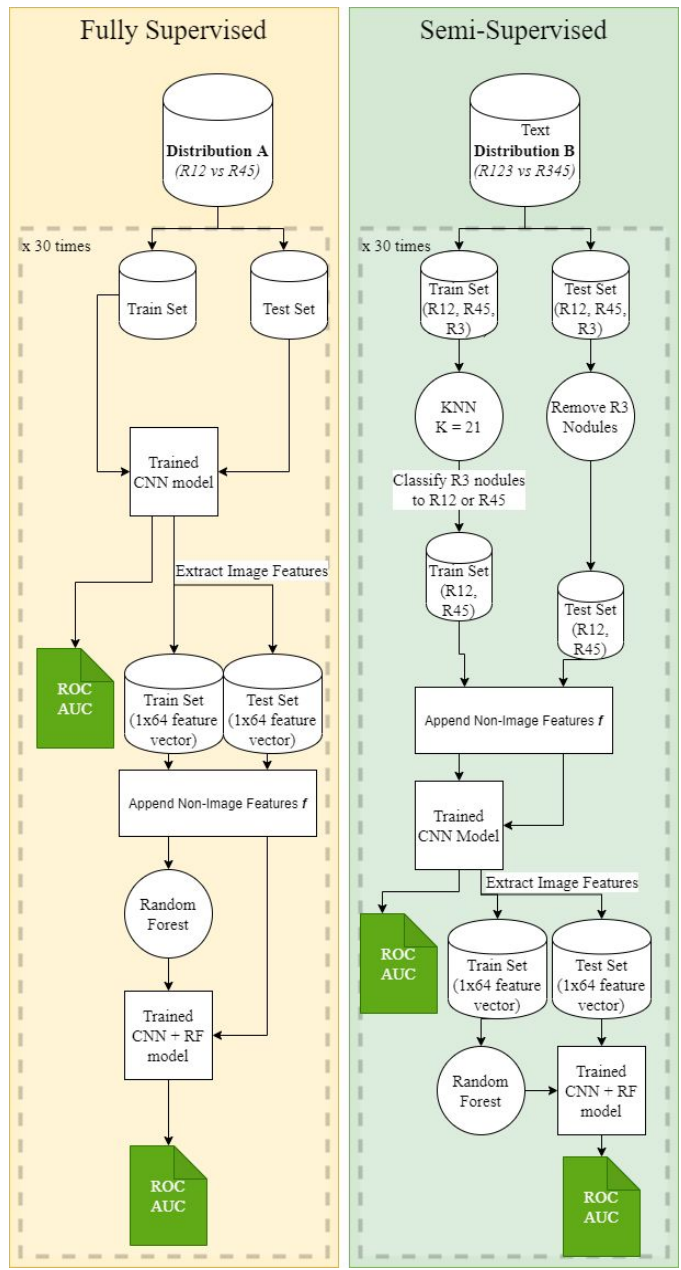

*Figure 7. General training and testing strategy for the supervised and semi-supervised models for experiments 3-5. The variable 'f' takes the value of the non-image features.*

C. *Experiment 3 – Classification using Images + Biomarkers*

A 3D CNN + Random Forest model is employed in this experiment. The 3D CNN model trained on the bounding boxes in experiment 2 is loaded on each of the 30 iterations here to extract the feature vectors on the loaded train and test set for the fully supervised method. We append the biomarker features to these image features for the corresponding rows in the train and test set. Since the image features have a length of 64 and the biomarkers have a length of 8, we repeat the biomarker features eight times before combining these features. Finally, we have a combined feature vector for a nodule of length 128 (64 image features + [8 biomarker features × 8]). A random forest is trained on these features, and the ROC AUC is calculated on the test set that contains the same features.

We follow the same suit of feature augmentation, training, and testing when the 3D CNN + Random forest model is trained using the semi-supervised approach. We use the same KNN classification techniques for R3 nodules in the train set and remove the R3 nodules from the test set. ROC AUC values for each iteration are recorded, and the average is calculated. As illustrated in **Figure 7**, the variable *f* is Biomarker features.

*D. Experiment 4 - Classification using Images + Volumetric Radiomic features*

This experiment is similar to experiment 3, but the only difference is that instead of appending the biomarkers to the image feature vector, we append the maximum diameter of the nodule in the axial plane, the surface area of the nodule and the volume of the nodule. These features are extracted using the pylidc library developed by Hancock et. al [2]. Since we have an image feature vector length of 64 and the radiomics feature vector length of 3, we append these features 21 times to the image feature vector to get a single feature vector of length 127. Just as in experiment 3, we train a random forest on the feature vector dataset and calculate the ROC AUC and plot the ROC curve as well as it's average ROC AUC. As illustrated in **Figure 7**, the variable *f* is Volumetric Radiomics.

*E. Experiment 5 - Classification using Images + Biomarkers + Volumetric Radiomics*

This experiment also follows the same strategy of training as in experiment 3 and 4. The difference here is that we append both the biomarker and the radiomic features to the image feature vector. The biomarker and radiomic features together form a vector of length 11. This is appended 6 times so that our final feature vector will be of length 130. As illustrated in **Figure 7**, the variable *f* is the combination of biomarkers and volumetric radiomics.

VI. RESULTS

We describe the results experiment-wise in this section and finally compare all the results.

A. *Experiment 1 – Biomarkers only*

For the fully supervised models (R12 vs R45), we observe an average ROC AUC of 0.9165 for logistic regression and 0.9339 for random forest. For the semi-supervised models (R123 vs R345), we observe an average ROC AUC of 0.9200 for logistic regression and 0.9396 for random forest. The distribution of the AUC's for both these models are illustrated in **Figure 8** . The ROC curves comparing these models for both the distributions are illustrated in **Figure 9.**

The random forest model performs better than the logistic regression model irrespective of the method of supervision used for training. We also observe an improvement in both the models when the semi-supervised method is used. We run T-tests to test the statistical significance of the differences in model performance between random forest and logistic regression, as well as the difference in performance between fully supervised and semi-supervised techniques. Our first alternate hypothesis is that the random forest model has a higher average AUC compared to the logistic regression model irrespective of the method of supervision. The P-value we get is <.0001 and this proves that the random forest is statistically significant when it comes to lowering the false positive rates while maintaining high true positive rates compared to a logistic regression model. Our second alternate hypothesis is that the semi-supervised implementation of both the RF and LR models has a higher average AUC compared to their fully-supervised counterparts. Once again, our T-test results support this hypothesis. The P-value observed is <.0001. The best performing model is that semi-supervised Random Forest model with an average AUC of 0.9396.

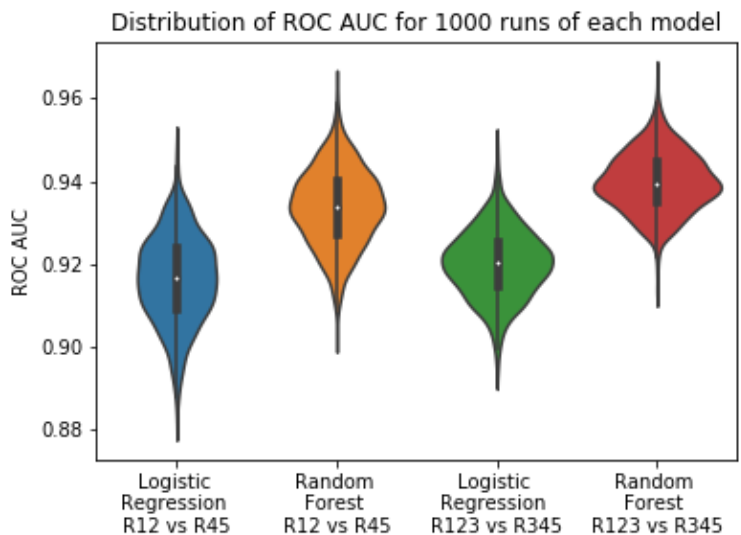

*Figure 8. – The violin plots illustrate the distribution of the logistic regression model and the random forest model over 1000 training sessions. Every training step uses a unique set of points in the 80:20 training and testing split*

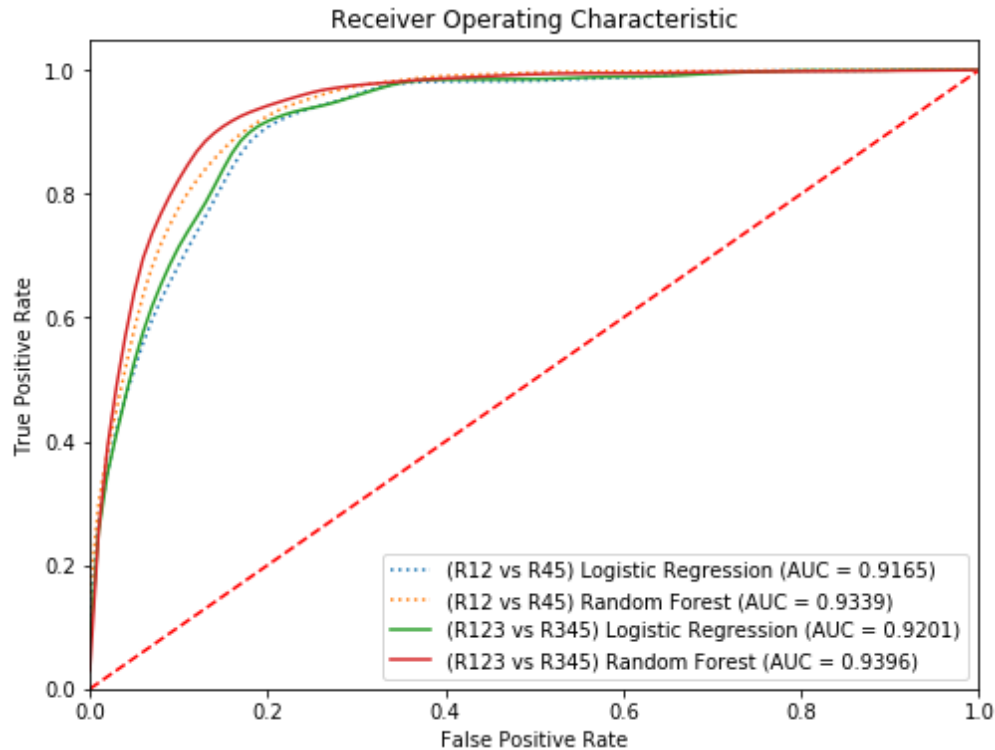

*Figure 9. – The Average ROC Curves for experiment 1 are plotted here. The ROC curves for fully-supervised models are plotted as a dotted line and the ROC curves for semi-supervised models are plotted as a solid line.*

B.  *Experiment 2 – Classification using images only*

The average ROC AUC we observe for the fully supervised 3D CNN is 0.7894. The average AUC increases to 0.7927 when a CNN + Random Forest classifier is used. For the semi-supervised models, the 3D CNN classifier achieves an average AUC of 0.8004 and the CNN + Random Forest classifier achieves an average AUC of 0.8072. We notice a higher AUC value for the semi-supervised models compared to the fully supervised ones. On performing T-tests to establish the statistical significance of this observation, we get a P-value <.05. The semi-supervised models perform better than the fully-supervised models. However, the addition of the random forest to the 3D CNN does not have any statistical significance when only image features are used for classification. We observe a change in this behaviour in experiments 3, 4 and 5.

C.  *Experiment 3 – Classification using Images + Biomarkers*

The average ROC AUC on the fully supervised 3D CNN + Random Forest model is 0.8652. This is a significant improvement compared to the AUCs in experiment 2. The AUC on the semi-supervised 3D CNN + Random Forest is 0.8647. These results are statistically significant when compared to the models that use only the images for the classification. However, the method of supervision does not seem to have a significant effect on how the models perform in this case.

*D. Experiment 4 - Classification using Images + Volumetric Radiomic Features*

The average ROC AUC on the fully supervised 3D CNN + Random Forest model is 0.8194 and the AUC on Semi-supervised 3D CNN + Random Forest model is 0.8361. Here, the semi-supervised model performs statistically better than the fully supervised model. The P-value after performing the T-test is <.0001.

*E. Experiment 5 - Classification using Images + Biomarkers + Volumetric Radiomic Features*

The average ROC AUC on the fully supervised 3D CNN + Random Forest model is 0.8659 and the AUC on the semi-supervised 3D CNN + Random Forest model is 0.8674.

**Figure 10** illustrates the distribution of the AUC values for each of the 30 runs for experiments 2, 3, 4 and 5 for the fully supervised models (R12 vs R45). Similarly, **Figure 12** illustrates the distribution of the AUC values for the semi-supervised models (R123 vs R345). **Figure 11** and **Figure 13** are plots of the ROC curve for experiments 2-5 for fully supervised and semi-supervised models respectively. **Table 1** summarizes the ROC AUC results for all the experiments. We discuss these results in the following subsection.

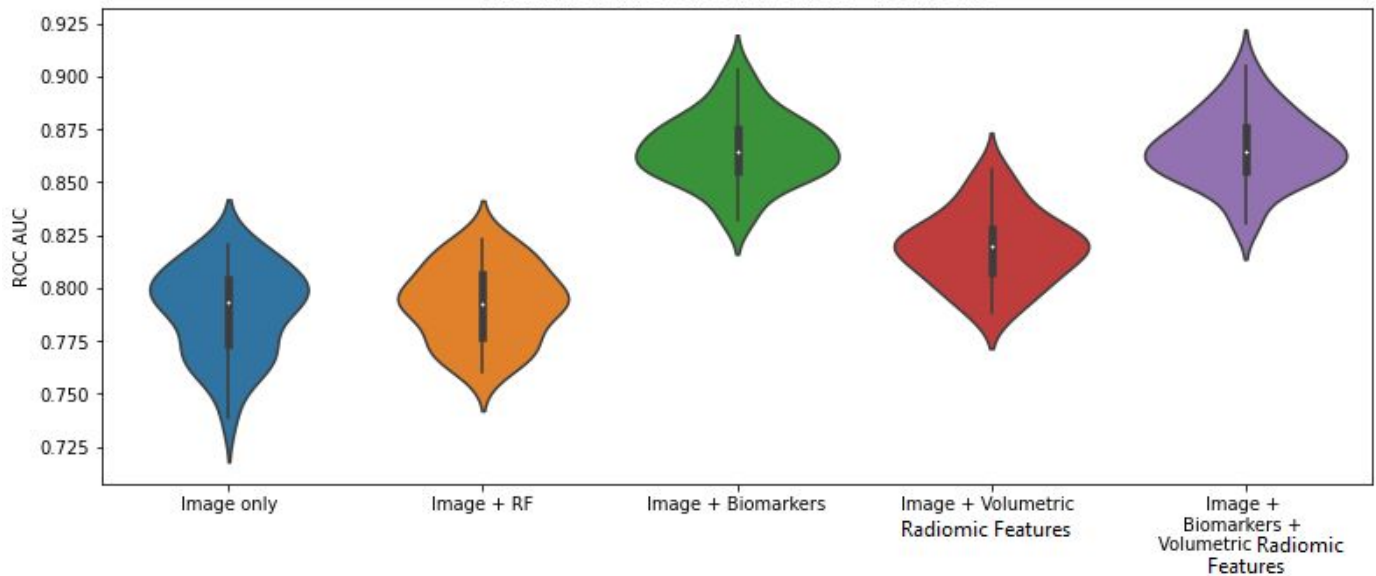

*Figure 10. The violin plots represent the distribution of the AUC scores for each of the experiments over 30 iterations of each experiment. We see that the AUC scores for the model that uses Image + Biomarkers in training and the model that uses Image + Biomarkers + Volumetri Radiomic features have competitive AUC distributions. This plot is for the fully supervised models.*

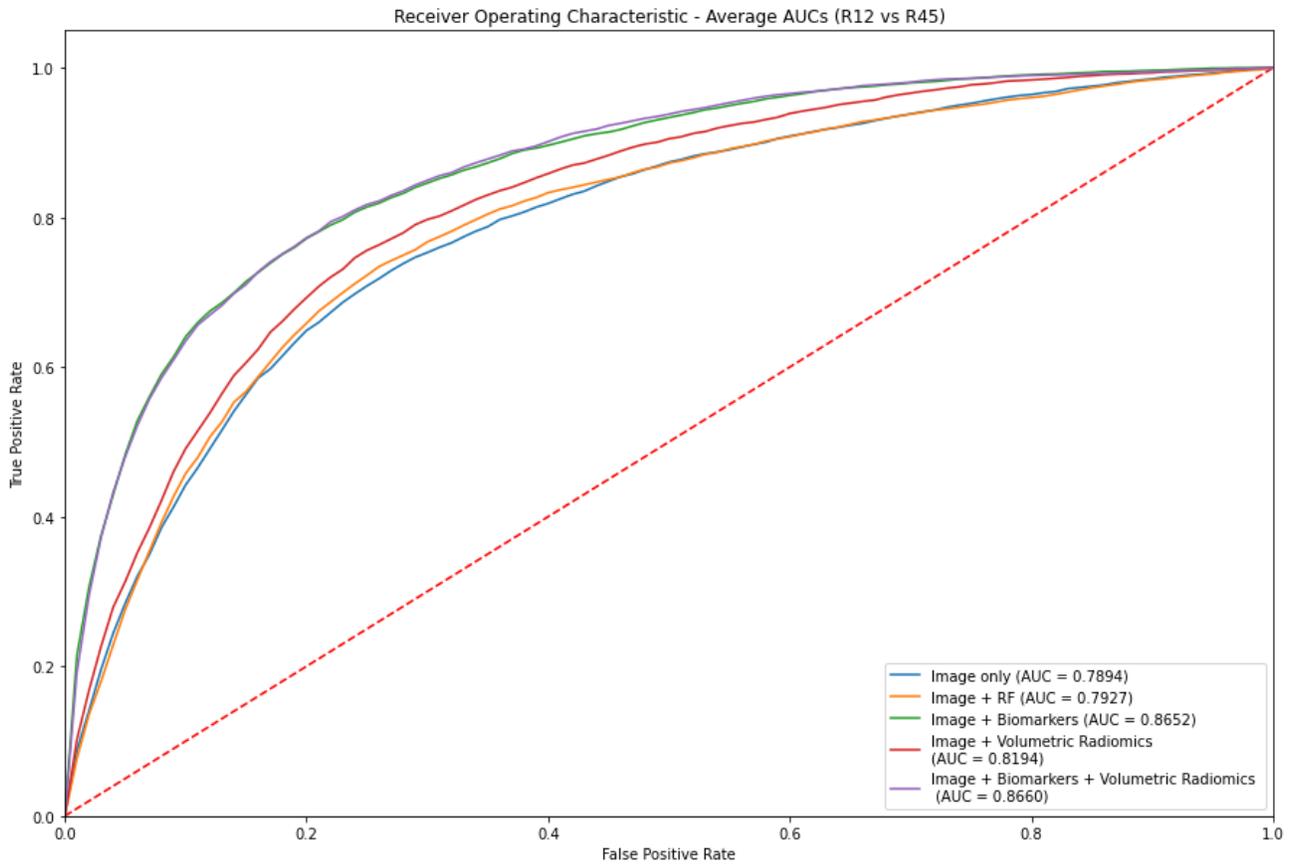

*Figure 11. The ROC curves for experiments 2-5 for fully supervised models.*

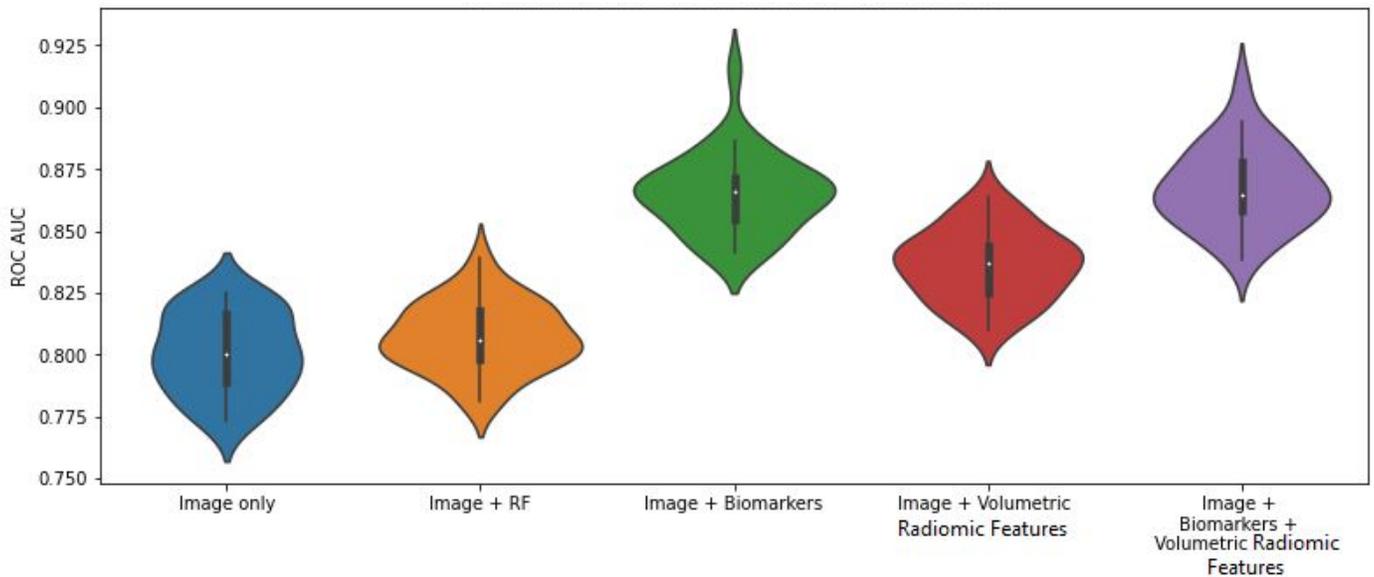

*Figure 12. The violin plots represent the distribution of the AUC scores for each of the experiments over 30 iterations of each experiment. This plot is for the semi-supervised models.*

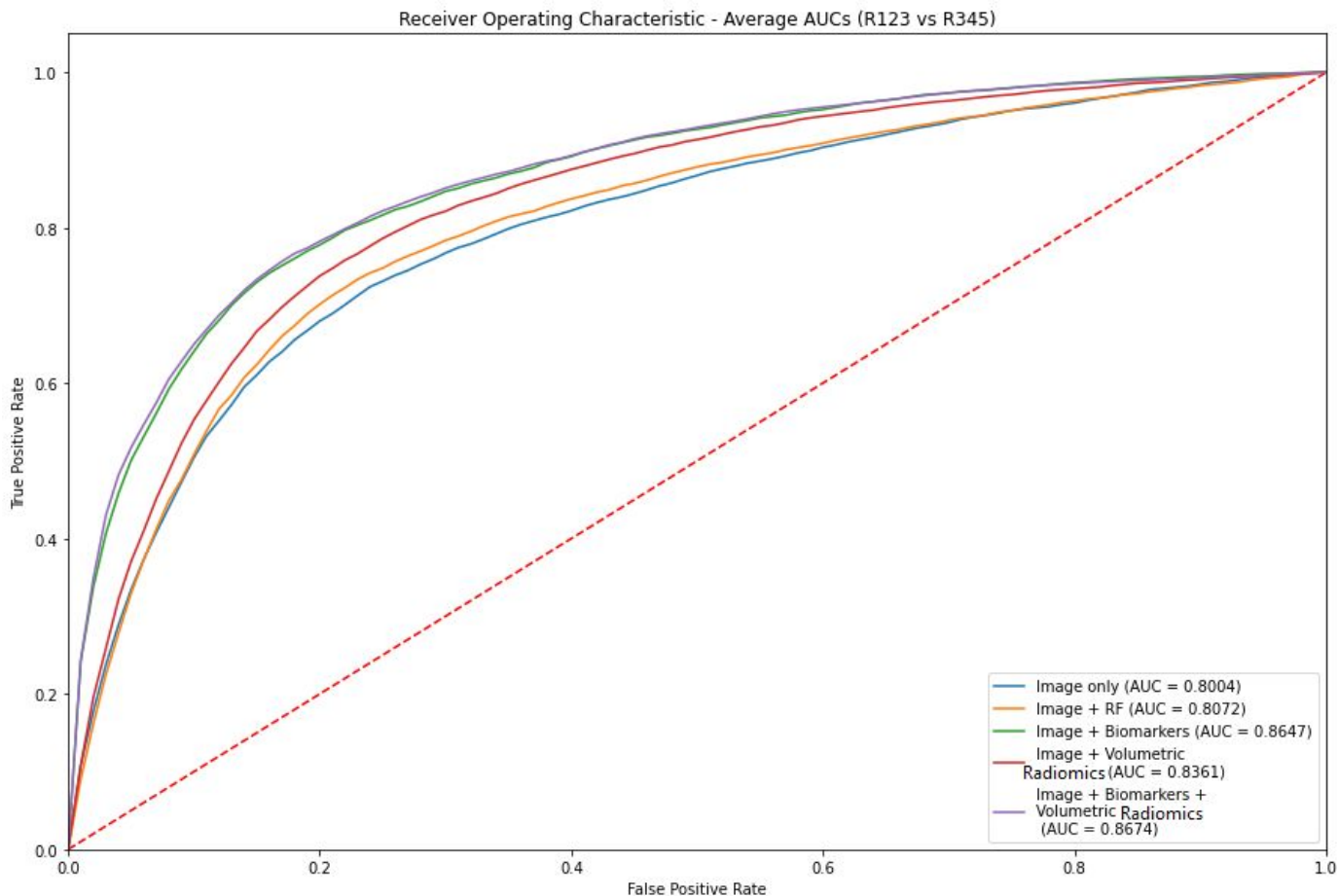

*Figure 13. The ROC curves for experiments 2-5 for semi-supervised models (R123 vs. R345)*

| Experiments | Models | Fully Supervised (R12 vs R45) ROC AUC | Semi-Supervised (R123 vs R345) ROC AUC |
|---|---|---|---|
| 1 – Classification using Biomarkers only | Logistic Regression | 0.9165 | 0.9201 |
| | Random Forest | 0.9339 | 0.9396 |
| 2 – Classification using Images only | CNN | 0.7894 | 0.8004 |
| | CNN + Random Forest | 0.7927 | 0.8072 |
| 3 – Classification using Images + Biomarkers | CNN + Random Forest (Biomarkers) | 0.8651 | 0.8646 |
| 4 - Classification using Images + Volumetric Radiomics | CNN + Random Forest (Radiomic features) | 0.8194 | 0.8361 |
| 5 - Classification using Images + Biomarkers + Volumetric Radiomics | CNN + Random Forest (Combined biomarker and radiomic features) | 0.8659 | 0.8674 |

*Table 1. Results for all experiments. Rows highlighted in yellow show best performing models that have statistical improvement compared to other models in their respective experiments. Rows in blue are not statistically different from each other.*

*F. Observation*

We observe that when only biomarkers are used, a non-linear model like the random forest is significantly better compared to the linear model. We also observe that the semi-supervised implementations of the models in all the experiments improve the AUC to a certain degree; specifically when models use biomarkers, images only and when models use the combination of images and volumetric radiomics. This behavior can be attributed to the underlying meaningful relationship between the features of the nodules to the malignancy values. By quantifying this relationship using semi-supervised methods, we can increase the amount of data the models can train on. This is a novel approach in the field of lung nodule malignancy classification as per our findings. To clearly observe the best performing combination of features, we must present the best performing models in the individual experiments in descending order of their average AUC values and perform t-tests between the models to prove the statistical significance of that order. On performing these tests, we summarize the results in **Table 2** and discuss them.

| Features | Model | Average AUC |
|---|---|---|
| Biomarkers only | Semi-Supervised Random Forest | 0.9396 |
| Images + Biomarkers + Volumetric features & Images + Biomarkers | Semi-supervised 3D CNN + RF and fully supervised 3D CNN + RF | ~0.8658 |
| Images + Volumetric features | Semi-Supervised CNN + RF | 0.8361 |
| Images only | Semi-Supervised CNN + RF | 0.8072 |

*Table 2. Best performing models that are statistically significantly rated in descending order of their average AUC values. Models that are grouped in one cell show that there is no statistical difference between them but both of them independently are better performing that the model in the row below them.*

Most existing methods make use of only one class of features (Biomarkers / Volumetric Radiomics / Image Features) for pulmonary nodule malignancy estimation. Our results show that the relationship between the features used to categorize the nodule malignancy suspicion can be made more meaningful when they are combined. The use of a combination of non-image features using the semi-supervised technique gives the best overall performance. The popular deep learning classification techniques that rely on images only can be significantly improved when non-image features like biomarkers and volumetric features are combined with them.

Nodules that may be classified as inconclusive by a human radiologist often have features that are similar to nodules that are either malignant or benign. Using semi-supervised methods, our algorithm can provide a second opinion to the radiologist about these nodules and has the potential to reduce the false positive rates.

VII. CONCLUSION

The aim of this research is to develop a hybrid lung nodule malignancy suspicion classification algorithm and effectively reduce the false-positive rates. Features that radiologists use to provide an opinion about the nodule malignancy are biomarkers and radiomic features. CADx systems are being used by human annotators to reduce the false-positive rates and provide a second opinion to the nodule classifications. Traditional methods for the computed aided diagnosis systems make use of a specific class of features for classification. These are either image-based, biomarkers, or radiomics. CNN's have been powerful when working with images. However, they require a large amount of data to learn the representation of the object. Conventional methods that use non-image features perform well for many tasks with relatively small sample sizes, but they are not able to fully reflect the unique characteristics of the nodule.

We approach this problem by combining the image-based features with biomarkers and volumetric radiomic features. Additionally, we use a semi-supervised method that includes nodules that have an intermediate malignancy label. As per our knowledge, this approach has not been explored and is novel to the lung nodule classification field.

We investigated four models – Logistic regression, Random Forest, 3D CNN, and CNN + Random Forest, with three feature combinations that include the supervised and semi-supervised implementations. Our results show that the combination of features gives a better AUC score compared to models that use image features alone. Semi-supervised implementations of these models improve the achieved AUC values. This effect is attributed to the increase in the meaningful training data and establishing meaningful relationships between the combination of features and malignancy.

An unexpected result of our analysis was that the random forest model using biomarkers only outperformed the proposed hybrid model combining biomarkers, 3D CNNs, volumetric radiomics and semi-supervised learning. Although this result would suggest that descriptive biomarkers alone are a superior feature for malignancy estimation, it is not clear to what extent this result might be influenced by cognitive bias as the malignancy estimates and biomarkers were ultimately recorded by the same radiologist panel and may be influenced by one another [14]. In future work we wish to further investigate potential sources of bias in the LIDC-IDRI through analysis of the 96 patients with additional pathology information.

VIII. DISCUSSION AND FUTURE WORK

We believe our results are competitive with the current state of the art models for similar tasks of nodule level malignancy estimation. However, it is important to note that minor differences in the evaluation criteria of different investigators using LIDC-IDRI for malignancy estimation make it difficult to directly compare the AUC of one published study versus another to determine state of the art. We can achieve significant improvement when compared to the models that use CNN's with images only as their features [3, 4, 5, 7]. However, certain studies like Causey et al. [12] report an AUC of 0.99 when radiomic and deep image features are combined although this AUC is for the task of patient-level malignancy, not a nodule level malignancy and thus not directly comparable. Our evaluation metrics are most similar to those of Hancock and Magnan [2] that report AUC of 0.916 using Biomarkers only and 0.932 combining Biomarkers + Volumetric features; these results compare to our AUC of 0.9396 for Biomarkers only, 0.9555 for Biomarkers + Volumetric features.

A potential challenge with the use of the LIDC-IDRI is that the malignancy estimates provided by a four radiologist panel are subject to information bias due to variations in clinical processes and expertise. Furthermore, as the malignancy estimates are created by the same panel of radiologists that record biomarkers, there is a possibility of cognitive bias in the event that the identification of descriptive biomarkers were influenced by the overall malignancy estimates [14]. Despite these potential sources of bias, LIDC-IDRI is a highly valuable resource used by many investigators to better predict malignancy suspicion levels using descriptive biomarkers, radiomic, deep image features, and combinations thereof [17]. A promising approach to potentially reduce this bias was investigated by Kang et. al [16] that attempted to correlate malignancy suspicion with pathology levels using a subset of 157 patients with pathology data. However, of these 157 patients, only 96 have both pathology data as well as malignancy suspicion levels available simultaneously. Furthermore, the pathology and malignancy estimates use different nodule identifiers thereby making it difficult to identify linking variables. In future work, we intend to incorporate the approach of Kang et al. [16] in order to correlate our results to the subset of the LIDC-IDRI dataset for which both malignancy suspicion and pathology data are available, thereby decoupling the biomarkers from malignancy estimates of LIDC-IDRI over these 96 patients. In future work we would also like to evaluate the influence of other semi-supervised learning techniques in place of KNN. The use of more sophisticated classifiers that group the inconclusive nodules into pseudo-labels based on all of the available features rather than just the biomarkers for the semi-supervised step has the potential to further improve the accuracy of this step.

State of the art CNN models like Alexnet, VGG-Net, and LeNet can be modified to accept 3D images. These models can be used in place of the ten layer CNN to improve the CNN component of our work. Additionally, the spatial locality feature of the nodule is shown to provide discriminating information when it comes to the malignancy classification task. Global views or multi-view CNN's have the potential to improve the overall image classification task. Combining biomarker and radiomic features with this has the potential to reduce the false-positive rates further.

IX. ACKNOWLEDGEMENTS

We would like to thank Dr. Eliot Seigel, Dr. Michael Morris, Dr. Yelena Yesh and the members of the VIPAR Lab and CARTA lab, UMBC for all the support, advice and valuable feedback for this research.

X. REFERENCES

[1]     Y. Liu, Y. Balagurunathan, T. Atwater, S. Antic, Q. Li, R. C. Walker, G. T. Smith, P. P. Massion, M. B. Schabath and R. J. Gillies, "Radiological image traits predictive of cancer status in pulmonary nodules," Clinical Cancer Research, vol. 23, no. 6, pp. 1442-1449, 15 3 2017.

[2]     M. C. Hancock and J. F. Magnan, "Lung nodule malignancy classification using only radiologist-quantified image features as inputs to statistical learning algorithms: probing the Lung Image Database Consortium dataset with two statistical learning methods.," Journal of medical imaging (Bellingham, Wash.), vol. 3, no. 4, p. 044504, 10 2016.

[3]     I. Bonavita, X. Rafael-Palou, M. Ceresa, G. Piella, V. Ribas and M. A. González Ballester, "Integration of convolutional neural networks for pulmonary nodule malignancy assessment in a lung cancer classification pipeline," Computer Methods and Programs in Biomedicine, vol. 185, p. 105172, 1 3 2020.

[4]     J. Kuruvilla and K. Gunavathi, "Lung cancer classification using neural networks for CT images," Computer Methods and Programs in Biomedicine, vol. 113, no. 1, pp. 202-209, 1 1 2014.

[5]     X. Zhao, L. Liu, S. Qi, Y. Teng, J. Li and W. Qian, "Agile convolutional neural network for pulmonary nodule classification using CT images," International Journal of Computer Assisted Radiology and Surgery, vol. 13, no. 4, pp. 585-595, 1 4 2018.

[6]     D. Kumar, A. Wong and D. A. Clausi, "Lung Nodule Classification Using Deep Features in CT Images," in Proceedings -2015 12th Conference on Computer and Robot Vision, CRV 2015, 2015.

[7]     W. Li, P. Cao, D. Zhao and J. Wang, "Pulmonary Nodule Classification with Deep Convolutional Neural Networks on Computed Tomography Images.," Computational and mathematical methods in medicine, vol. 2016, p. 6215085, 2016.

[8]     Y. LeCun, L. Bottou, Y. Bengio and P. Haffner, "Gradient-based learning applied to document recognition," Proceedings of the IEEE, vol. 86, no. 11, pp. 2278-2323, 1998.

[9]     A. Krizhevsky, I. Sutskever and G. E. Hinton, "ImageNet Classification with Deep Convolutional Neural Networks".

[10]    F. Liao, M. Liang, Z. Li, X. Hu and S. Song, "Evaluate the Malignancy of Pulmonary Nodules Using the 3-D Deep Leaky Noisy-OR Network," IEEE Transactions on Neural Networks and Learning Systems, vol. 30, no. 11, pp. 3484-3495, 1 11 2019.

[11]    M. B. Rodrigues, R. V. M. Da Nobrega, S. S. A. Alves, P. P. R. Filho, J. B. F. Duarte, A. K. Sangaiah and V. H. C. De Albuquerque, "Health of Things Algorithms for Malignancy Level Classification of Lung Nodules," IEEE Access, vol. 6, pp. 18592-18601, 22 3 2018.

[12]    J. L. Causey, J. Zhang, S. Ma, B. Jiang, J. A. Qualls, D. G. Politte, F. Prior, S. Zhang and X. Huang, "Highly accurate model for prediction of lung nodule malignancy with CT scans," Scientific Reports, vol. 8, no. 1, 1 12 2018.

[13]    S. Li, P. Xu, B. Li, L. Chen, Z. Zhou, H. Hao, Y. Duan, M. Folkert, J. Ma, S. Huang, S. Jiang and J. Wang, "Predicting Lung Nodule Malignancies by Combining Deep Convolutional Neural Network and Handcrafted Features".

[14]    S. G. Armato, G. McLennan, L. Bidaut, M. F. McNitt-Gray, C. R. Meyer, A. P. Reeves, B. Zhao, D. R. Aberle, C. I. Henschke, E. A. Hoffman, E. A. Kazerooni, H. MacMahon, E. J. Van Beek, D. Yankelevitz, A. M. Biancardi, P. H. Bland, M. S. Brown, R. M. Engelmann, G. E. Laderach, D. Max, R. C. Pais, D. P. Qing, R. Y. Roberts, A. R. Smith, A. Starkey, P. Batra, P. Caligiuri, A. Farooqi, G. W. Gladish, C. M. Jude, R. F. Munden, I. Petkovska, L. E. Quint, L. H. Schwartz, B. Sundaram, L. E. Dodd, C. Fenimore, D. Gur, N. Petrick, J. Freymann, J. Kirby, B. Hughes, A. Vande Casteele, S. Gupte, M. Sallam, M. D. Heath, M. H. Kuhn, E. Dharaiya, R. Burns, D. S. Fryd, M. Salganicoff, V. Anand, U. Shreter, S. Vastagh, B. Y. Croft and L. P. Clarke, "The Lung Image Database Consortium (LIDC) and Image Database Resource Initiative (IDRI): A completed reference database of lung nodules on CT scans," Medical Physics, vol. 38, no. 2, pp. 915-931, 2011.

[15]    V. Parekh and M. A. Jacobs, Radiomics: a new application from established techniques, vol. 1, Taylor and Francis Ltd., 2016, pp. 207-226.

[16] G. Kang, K. Liu, B. Hou and N. Zhang, "3D multi-view convolutional neural networks for lung nodule classification," PLOS ONE, vol. 12, no. 11, p. e0188290, 16 11 2017.

[17] X. Wang, K. Mao, L. Wang, P. Yang, D. Lu and P. He, An appraisal of lung nodules automatic classification algorithms for CT images, vol. 19, MDPI AG, 2019.

## XI. APPENDIX

The random forest model is tuned for each of the experiments to obtain optimal classification accuracy. For this we use a random search cross validation approach. We define a grid of hyperparameter ranges for the maximum number of features considered for splitting a node, number of trees in the forest, maximum number of levels in each decision tree, minimum number of data points placed in the node before the node is split, minimum number of data points allowed in a leaf node and the method of sampling data points. We use a 3-fold cross validation method with every random sample of parameters. This process is iterated 100 times for each fold and the model is compared with all the possible outcomes. The parameters of the best performing model are then extracted for the training process in the experiments 1-5 for each dataset distribution. **Table 1 and 2** consist of the random forest hyperparameters for each of the experiments.

Table 1 - Random Forest hyperparameters for experiments 1-5 for data distribution R12 vs R45

| Experiment | Number of estimators | Maximum depth | Maximum features | Minimum samples per leaf | Minimum samples per split |
|---|---|---|---|---|---|
| Exp 1: Biomarkers only | 1000 | 10 | square root of number of features | 1 | 5 |
| Exp 2: Images only | 200 | 10 | square root of number of features | 4 | 2 |
| Exp 3: Images + Biomarkers | 800 | 10 | square root of number of features | 2 | 5 |
| Exp 4: Images + Volumetric Radiomics | 400 | 100 | square root of number of features | 4 | 10 |
| Exp 5: Images + Biomarkers + Volumetric Radiomics | 300 | 110 | square root of number of features | 2 | 2 |

Table 2 - Random Forest hyperparameters for experiments 1-5 for data distribution R123 vs R345

| Experiment | Number of estimators | Maximum depth | Maximum features | Minimum samples per leaf | Minimum samples per split |
|---|---|---|---|---|---|
| Exp 1: Biomarkers only | 200 | 100 | square root of number of features | 1 | 5 |
| Exp 2: Images only | 200 | 10 | square root of number of features | 4 | 2 |
| Exp 3: Images + Biomarkers | 600 | 10 | square root of number of features | 2 | 5 |
| Exp 4: Images + Volumetric Radiomics | 200 | 10 | square root of number of features | 4 | 2 |
| Exp 5: Images + Biomarkers + Volumetric Radiomics | 500 | 110 | square root of number of features | 1 | 2 |